# Improving indistinguishability of single photons from colloidal quantum dots using nanocavities

Abhi Saxena[1], Yueyang Chen[1], Albert Ryou[1], Carlos G. Sevilla[3], Peipeng Xu[1,4,5], Arka Majumdar[1,2,*]

[1]Electrical and Computer Engineering, University of Washington, Seattle, Washington 98195, United States

[2]Department of Physics, University of Washington, Seattle, Washington 98195, United States

[3]School of Natural Science, Hampshire College, Amherst, MA 01002, United States

[4]Laboratory of Infrared Materials and Devices, Advanced Technology Research Institute, Ningbo University, Ningbo 315211, China

[5]Key Laboratory of Photoelectric Detection Materials and Devices of Zhejiang Province, Ningbo, 315211, China

[*]Corresponding Author: arka@uw.edu

KEYWORDS: Indistinguishable single photon source, Colloidal quantum dots

Abstract:

Colloidal quantum dots have garnered active research interest as quantum emitters due to their robust synthesis process and straightforward integration with nanophotonic platforms. However, obtaining indistinguishable photons from the colloidal quantum dots at room temperature is fundamentally challenging because they suffer from an extremely large dephasing rate. Here we propose an experimentally feasible method of obtaining indistinguishable single photons from an incoherently pumped solution-processed colloidal quantum dot coupled to a system of nanocavities. We show that by coupling a colloidal quantum dot to a pair of silicon nitride cavities, we can obtain comparable performance of a single photon source from colloidal quantum dots as other leading quantum emitters like defect centers and self-assembled quantum dots.

**Introduction**

Hybrid quantum photonic integrated circuits[1,2] are a promising platform to develop various quantum technologies including universal quantum computing[3,4], quantum networks[5] and boson sampling[6]. A fundamental building block of this hybrid quantum photonic platform is an on-chip source of indistinguishable single photons. Quantum emitters including self-assembled quantum dots (QDs) and single defect centers coupled to integrated nanocavities have recently attracted significant attention as indistinguishable single photon sources due to their on-demand and high-rate single photon generation capabilities[7,8]. The indistinguishability of these solid-state emitters which is largely limited by dephasing, is mitigated by using an optical cavity in these systems[9]. Unfortunately, none of these has been shown to maintain indistinguishability of generated single photons on a scalable platform which is a prerequisite for most quantum technologies. Solution-processed colloidal QDs can potentially provide a promising solution to this problem due to their low-cost chemical synthesis and straightforward deposition to most substrates in a scalable manner. In fact, deterministic positioning of colloidal QDs on silicon nitride (SiN) integrated photonic platform has been recently demonstrated[10]. However, despite the ease of scalable fabrication, solution-processed colloidal QDs suffer from a large dephasing rate ($\gamma^* \approx 10^5 \gamma$, $\gamma^*$ being the pure dephasing rate and $\gamma$ being the QD dipole decay rate) at room temperature making them unattractive as an indistinguishable single photon source.

In this paper we report an architecture consisting of a colloidal QD coupled to two nanophotonic resonators that improves the indistinguishability of single photons while maintaining a moderate efficiency at room temperature. Specifically, we show that our architecture can achieve comparable efficiency and indistinguishability of single photons from colloidal QDs even though they suffer from order of magnitude greater dephasing than quantum emitters like silicon vacancy (SiV) centers[11]. We theoretically analyze the parameter space of our nanophotonic architecture to identify regions of high indistinguishability and efficiency. Finally, we propose an experimentally viable system to implement the architecture under the constraints of current nanofabrication technology required to obtain indistinguishable single photons from colloidal QDs. We note that our proposed method is inspired by recent work on improving indistinguishability of single photons emitted by SiV center using cascaded cavities[11].

**Indistinguishable photons from broad dissipative emitters**

For quantum emitters with large dephasing rate, the indistinguishability $I$ of emitted photons is given by[12,13]

$$I = \frac{\gamma}{\gamma + \gamma^*}$$

where $\gamma$ is the radiative decay rate and $\gamma^*$ is the pure dephasing rate of the quantum emitter. For solid and colloidal state quantum emitters at room temperatures, constantly varying local environmental conditions cause $\gamma^*$ to be much larger than $\gamma$[14,15]. This effect is particularly severe for colloidal QDs where $\gamma^* \approx 10^5 \gamma$ and the indistinguishability $I$ comes out to be $\sim 10^{-5}$, making it impossible to use the bare emitter as a useful indistinguishable single photon source.

For comparatively less dissipative emitters ($\gamma^* \lesssim 10^3 \gamma$) such as single self-assembled QDs or defect centers, regions of high indistinguishability based on different mitigating techniques have been theoretically identified including: cavity-funneling of indistinguishable photons in dielectric systems[13], usage of ultra-small mode volume cavity to boost indistinguishability primarily in plasmonic systems[16] or using a cascaded cavity system to get highly indistinguishable photons[11]. However, no reports exist for improving indistinguishability of emitted photons from strongly dissipative emitters like solution-processed colloidal QDs.

Our proposed system consists of two coupled cavities $C_1, C_2$ and a colloidal QD pumped with a picosecond pulse as shown in Figure 1(a). Cavity $C_1$ has a decay rate of $\kappa_1$ and is coupled to the emitter with coupling rate $g$. The second cavity $C_2$ decays at a rate of $\kappa_2$ and is coupled to $C_1$ with a coupling rate of $J$. The photons lost by $C_2$ are collected as the output of the system. We can see from Figure 1(b) that the linewidth of the quantum emitter $\gamma + \gamma^*$, is much broader than the linewidths of the cavities $\kappa_1, \kappa_2$ under consideration because of the huge dephasing experienced by the emitter.

Our system is governed by the Hamiltonian (setting $\hbar = 1$)

$$H = \omega_e e^\dagger e + \omega_{c_1} c_1^\dagger c_1 + \omega_{c_2} c_2^\dagger c_2 + g(e^\dagger c_1 + e c_1^\dagger) + J(c_1^\dagger c_2 + c_1 c_2^\dagger)$$

where $e^\dagger, c_1^\dagger, c_2^\dagger$ are the creation operators for the emitter and the cavities $C_1$ and $C_2$ respectively.

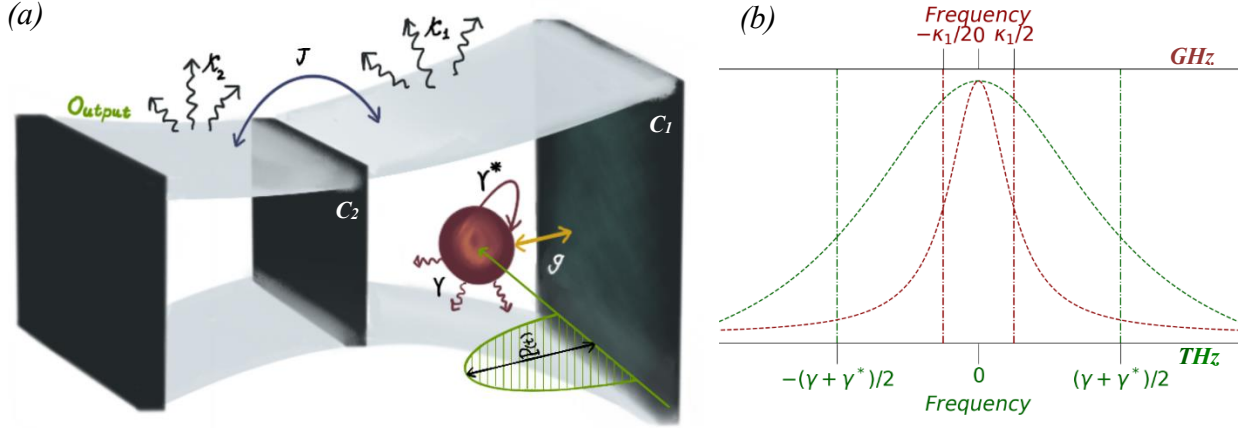

Figure 1: System description. (a) Quantum emitter with radiative decay rate $\gamma$ and pure dephasing rate $\gamma^*$ is coupled to an optical cavity $C_1$ with coupling rate $g$. The cavity has a decay rate of $\kappa_1$ and is coupled to another cavity $C_2$ with coupling rate $J$. The second cavity $C_2$ loses photons at a decay rate of $\kappa_2$ which are collected as the output of the system. The emitter is excited incoherently through a pump pulse of amplitude $P(t)$. (b) Superimposed spectra ($\omega_o = 0$) of the quantum emitter(green) and the optical cavity $C_1$(red) plotted on two differently scaled axes. Linewidth of emitter $\gamma + \gamma^* \gg \kappa_{1,2}$ the linewidths of the cavities in our system, $\kappa_1/2\pi = 7.9 GHz$, $\gamma/2\pi = 0.2 GHz$, $\gamma^*/2\pi = 17.4 THz$, $Q_1 = 6 \times 10^4$, $\omega_o/2\pi = 476 THz$.

The system dynamics is given by the evolution of the density matrix according to the master equation[17,18]

$$\frac{\partial \rho}{\partial t} = -i[H, \rho(t)] + \sum_n \left[\frac{1}{2}\left(2A_n \rho(t) A_n^\dagger - \rho(t) A_n^\dagger A_n - A_n^\dagger A_n \rho(t)\right)\right]$$

where $A_n$ denotes the collapse operators required to model the system: $\sqrt{\kappa_1} c_1, \sqrt{\kappa_2} c_2, \sqrt{\gamma} e, \sqrt{\gamma^*} e^\dagger e, \sqrt{P(t)} e^\dagger$ where the last term represents incoherent pumping of the QD. The collapse operator for incoherent pumping $\sqrt{P(t)} e^\dagger$ is time dependent to denote a gaussian pulse used to excite the emitter. $P(t)$ is given by

$$P(t) = P_o e^{\frac{-(t-t_o)^2}{2\sigma^2}}, \quad P_o = 120\gamma$$

where σ is the standard deviation corresponding to the width of the gaussian pulse centered at $t_o$. We emphasize that the previous works[11,13,16] modelled the single photon source by assuming an initially excited emitter, which is strictly valid only for resonant excitation. In most experiments, however, the single photons are generated under above-band pumping, and hence in our model we explicitly incorporated the incoherent pumping of the emitter using a pulsed laser. We note that the system Hamiltonian remains the same for both the incoherent pumping and the case of an initially excited emitter. The difference appears only in the collapse operators needed to model the system using the master equation.

The indistinguishability of photons emitted by the $C_2$ can calculated as[13]

$$I = \frac{\int_0^\infty dt \int_0^\infty d\tau \left|\langle c_2^\dagger(t+\tau)c_2(t)\rangle\right|^2}{\int_0^\infty dt \int_0^\infty d\tau \langle c_2^\dagger(t)c_2(t)\rangle\langle c_2^\dagger(t+\tau)c_2(t+\tau)\rangle}$$

We can define the efficiency $\beta$ of the system as the number of photons emitted by $C_2$ which is given by

$$\beta = \kappa_2 \int_0^\infty \langle c_2^\dagger(t)c_2(t)\rangle dt$$

We calculate both of these quantities via quantum regression theorem using QuTiP[19].

**Parameter study of indistinguishability and efficiency**

In this section we study how indistinguishability $I$ and efficiency $\beta$ vary with $Q_2$ (quality factor of $C_2$) and $J$ (coupling between the cavities $C_1$ and $C_2$), which are our primary degrees of freedom during the design process as well as being the key parameters in determining the system performance. We pump our emitter, a colloidal QD, with a $3\ ps$ gaussian pulse centered at $t_o =$ $5ps$ and calculate $I$ and $\beta$ as we move across the parameter space in order to identify regions of high indistinguishability and moderate efficiency.

To gain a physically intuitive understanding behind calculated values of $I$ and $\beta$ we study the population dynamics of the system. Qualitatively we can break down the sequence of events that generates single photons as follows (Figure 2). Before the incoherent excitation pulse hits the emitter, the emitter is in the ground state and the cavities are empty. When the pulse hits the emitter the population of the emitter rises at rate $P(t)$ while simultaneously decaying at rate $\gamma$ and getting dephased at a rate of $\gamma^*$. Meanwhile the cavity $C_1$ experiences a spike in its population, the magnitude of which depends on $g$, before its population decays back to zero with a decay rate of $\kappa_1$. The population in the second cavity $C_2$ also experiences a period of rise due to cavity coupling rate $J$, followed by an eventual decline to the ground state as the cavity emits photons and decays at rate $\kappa_2$. We collect these photons emitted by cavity $C_2$ and we want these to be as indistinguishable as possible while still being collected at practical collection efficiencies. By adiabatically eliminating the coherences in the optical Bloch equations describing the system (see Supplementary Material), we define $R_1$ as the bi-directional population transfer rate between emitter and first cavity $C_1$ and $R_2$ the bi-directional population transfer rate between the two cavities $C_1$ and $C_2$ (Figure 2). These are given by[11,13,20]

$$R_1 = \frac{4g^2}{\gamma + \gamma^* + \kappa_1}, R_2 = \frac{4J^2}{R_1 + \kappa_1 + \kappa_2}$$

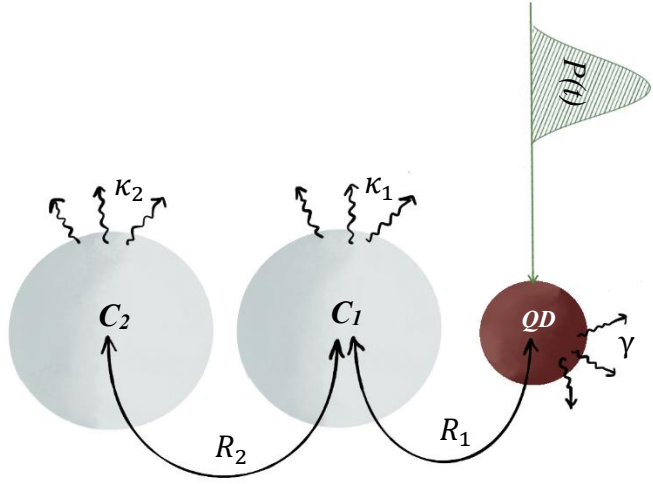

*Figure 2: System schematic for population dynamics. The colloidal QD which has a radiative decay rate $\gamma$ is pumped with an inchoerent pulse $P(t)$. The popluation transfer between the colloidal QD and $C_1$ occurs with a rate $R_1$. $C_1$ has a decay rate of $\kappa_1$. Population transfer rate between $C_1$ and $C_2$ is $R_2$. $C_2$ decays with a rate $\kappa_2$.*

In Figures 3(a), (b) we calculate $I$ and $\beta$ using the master equation and plot them as a function of $Q_2$ (or $\kappa_2 = \omega_o/Q_2$). We consider a system with $Q_1 = 6 \times 10^4$ and $J = 2.1\gamma$ and sweep across $Q_2$ for four different mode volumes ($V_{eff}$) of $C_1$. The coupling strength $g$ is inversely proportional to the $\sqrt{V_{eff}}$ whereas, $R_1$ varies monotonically with $g$. In Figure 3(c) we plot how $R_2$ changes as we sweep across $\kappa_2$.

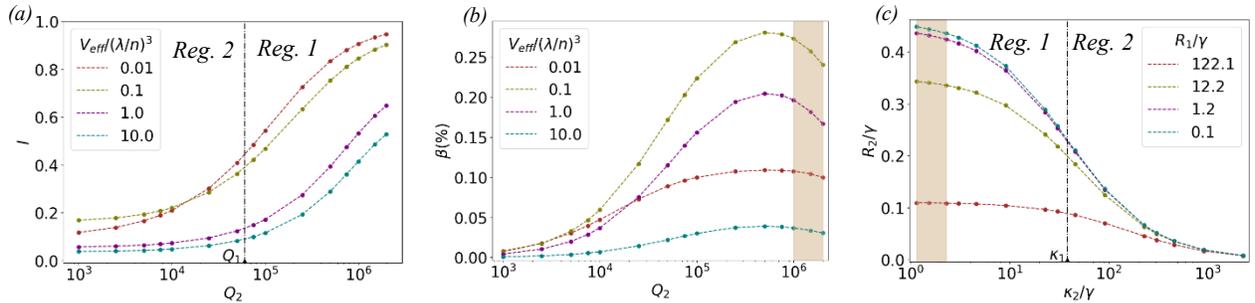

*Figure 3: Parameter study of indistinguishability $I$ and efficiency $\beta$. Each figure has four plots corresponding to four different mode volumes of the first cavity $C_1$ which corresponds to four different $g$'s and hence four different $R_1$'s. (a) $I$ as a function of $Q_2(\kappa_2)$. (b) $\beta$ as a function of*

$Q_2(\kappa_2)$. (c) $R_2$ as a function of $\kappa_2(Q_2)$. The figures (a), (c) are divided into two regions depending on the relative value of $\kappa_2(Q_2)$ with respect to $\kappa_1(Q_1)$. In Region 1. $\kappa_2 < \kappa_1$ and in Region 2. $\kappa_2 > \kappa_1$. The shaded areas in (b), (c) denote the region where $\kappa_2$ starts becoming comparable to $R_2$. Here, $Q_1 = 6 \times 10^4$, $J = 2.1\gamma$.

We divide the plots in Figure 3(a), (c) into two regions based on the relative value of $\kappa_2(Q_2)$ with respect to $\kappa_1(Q_1)$. In Region 1 where $\kappa_2 < \kappa_1$, i.e. the light storage time in the second cavity $C_2$ is larger than the first cavity $C_1$, the cavity $C_1$ gets populated with rate $R_1$ from the emitter, and acts like an emitter itself for the cavity $C_2$. $C_2$ then funnels the emission into its linewidth and consequently, we see a high $I$ which increases with a decreasing $\kappa_2$ or increasing $Q_2$. However, in Region 2 where $\kappa_2 > \kappa_1$ this funneling cannot happen efficiently and $I$ rapidly falls. As this process is boosted by an increased $R_1$, we expect $I$ to increase with an increasing $R_1$. This analysis is valid when there is a dominant unidirectional flow of single photons from QD towards $C_2$ (Figure 2), which dictates $R_2 \lesssim \kappa_2$ and $R_1 \lesssim \kappa_1 + R_2$. This ensures that photons do not incoherently hop back and forth between $C_1$ and $C_2$ or between the emitter and $C_1$ at rates which are higher than the rate of loss $\kappa_2$ from $C_2$.

As we increase $Q_2$ or decrease $\kappa_2$, we expect the efficiency $\beta$ to increase as we are increasing $R_2$ the rate of transfer to $C_2$ (Figure 3(b)). This indeed happens until a decreasing $\kappa_2$ starts becoming comparable to $R_2$ (shaded region in Figures 3(b), (c)), after which instead of photons being collected at output through $C_2$, they are returned to $C_1$ at a faster rate which lowers the $\beta$. This leads to the non-monotonic behavior of the efficiency as a function of $Q_2$. Next we look at effect of $R_1$ on $\beta$ as $Q_2$ is varied. For the system to emit with a non-zero $\beta$, $R_1$ must be greater than $\gamma$ because if $R_1 < \gamma$ most of photons will be lost by the emitter itself without being transferred to the cavities leading to extremely low $\beta$ values. Nonetheless, as we increase $R_1$, $R_2$ decreases and due to this inverse relationship, $\beta$ varies non-monotonically with $V_{eff}$ or $R_1$ (Figure 3(b)). A large enough $R_1$ is required to ensure that $C_1$ is populated so that a transfer to $C_2$ can happen. But an increased $R_1$ also leads to a lower $R_2$ and hence an efficiency maximum exits at an intermediate value of $R_1$ or correspondingly $V_{eff}$.

To get a better understanding of $I$ and $\beta$, in Figures 4(a), (b) we plot how $I$ and $\beta$ change with $J$, keeping $Q_1 = 6 \times 10^4, Q_2 = 2 \times 10^6$ constant for four different $V_{eff}$ or $g$ values. Value of $Q_2$ is chosen such to ensure we can reach high $I$ values as we vary $J$. In Figure 4(c) we plot how $R_2$ changes when we sweep across $J$ in Figures 4(a), (b).

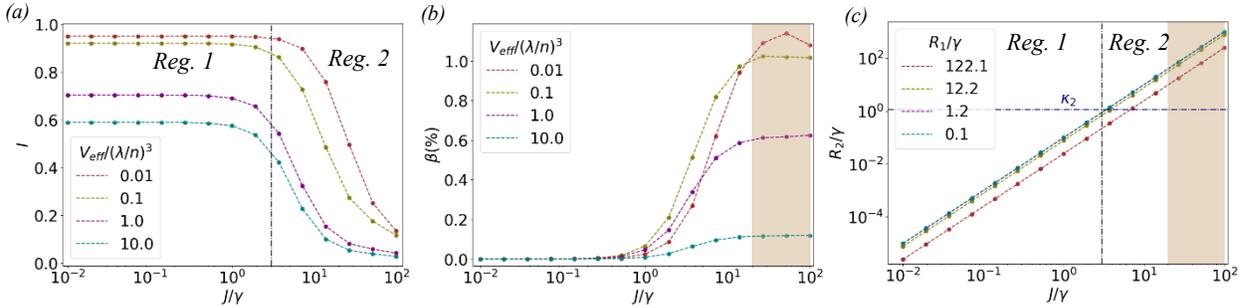

*Figure 4: Parameter study of indistinguishability I and efficiency β. Each figure has four plots corresponding to four different mode volumes of the first cavity $C_1$ which corresponds to four different $g_1$'s and hence four different $R_1$'s. (a) I as a function of J. (b) β as a function of J. (c) $R_2$ as a function of J. Here $Q_1 = 6 \times 10^4$, $Q_2 = 2 \times 10^6$. The figures (a), (c) are divided into two regions depending on the relative value of $R_2$ with respect to $\kappa_2$. In Region 1. $R_2 < \kappa_2$ and in Region 2. $R_2 > \kappa_2$. The shaded area in (b), (c) denotes the region where $R_2 \gtrsim R_1$.*

We first analyze the plots in Figure 4(a). With the help of Figure 4(c) we deconstruct the figure into two regions Region 1 where $R_2 < \kappa_2$ and Region 2 where $R_2 > \kappa_2$. In Region 1 where $R_2 < \kappa_2$, as we increase $J$, $R_2$ increases without having any significant effect on $I$. This happens because in this region our system is operating like the Region 1 described in Figure 3(a) with cavity $C_2$ funneling the acquired photons from $C_1$ into a narrow region as it emits them. As $\kappa_2$ is held constant $I$ doesn't change significantly. The curves are arranged in an order of increasing $R_1$ similarly. A significant change happens in the Region 2 where $R_2 > \kappa_2$. In this region $J$ has become significantly large to dominate the rate dynamics and this causes the photons to incoherently go back and forth between the two cavities at much higher rates than the emission rate of $C_2$ and this leads to a rapid decrease in $I$.

Finally, we consider Figure 4(b). Changing $J$ affects $R_2$ only, keeping $R_1$ unchanged. When $J < \gamma$, $R_2$ is very small and we find that $\beta$ is virtually zero because hardly any population transfer occurs between the cavities. As $J$ increases, $R_2$ also increases leading to an increase in $\beta$. In Figure 3(b) we observed that $\beta$ varied non-monotonically with $R_1$. However, now when $J$ becomes large enough that $R_2 \gtrsim R_1$ denoted by the shaded area in Figure 4(b), (c), we observe that an increasing $R_1$ leads to an increasing $\beta$. This is because now the limiting factor on $\beta$ is not the rate of transfer from $C_1$ to $C_2$ which is huge but rather $C_1$ getting populated in the first place. Note that despite high internal rates of transfer the overall magnitude of $\beta$ is still small because it is limited by $\kappa_2$ which is constant in this case and much smaller than $R_2$ when $J$ becomes large.

From this analysis we can find the optimal region of operation for our system. We need a high $Q_2$ and low $J$ to ensure that emitted photons from $C_2$ have high indistinguishability. This, however, restricts us to operate in the region of low efficiency. There exists a tradeoff between the indistinguishability and the efficiency in this region with the optimal point of operation at: $J$ just larger than $\gamma$, where indistinguishability has not dropped significantly but the efficiency too rises to a moderate value. Further in this region we can maximize the efficiency by choosing $V_{eff}$ between $0.1 \left(\frac{\lambda}{n}\right)^3$ to $1 \left(\frac{\lambda}{n}\right)^3$ with lower mode volumes preferred as they lead to a higher indistinguishability.

**Experimental Design**

Based on the parameter studies of the previous section we propose a SiN based nanophotonic structure achievable with current fabrication and experimental techniques to improve the indistinguishability of emitted photons from colloidal QDs. We observed in the last section that a mode volume of the first cavity between $0.1\left(\frac{\lambda}{n}\right)^3$ to $1\left(\frac{\lambda}{n}\right)^3$ is required to achieve optimal performance from our system. Based on the current state of art, on-substrate SiN cavities can achieve a mode volume of $\sim 1.2(\lambda/n)^3$ with a Q factor of $\sim 6 \times 10^4$ in a one-dimensional nanobeam structure (see Supplementary Material). For the second cavity we want a large $Q_2$, while the mode volume is not important. Hence, we can employ SiN ring resonator with a quality factor $Q_2 = 2 \times 10^6$ for this purpose[21,22]. Finally, we want $J$ to be slightly larger than $\gamma$ and hence we choose $J = 2.1\gamma$. This can be engineered by appropriately choosing the distance between the two cavities.

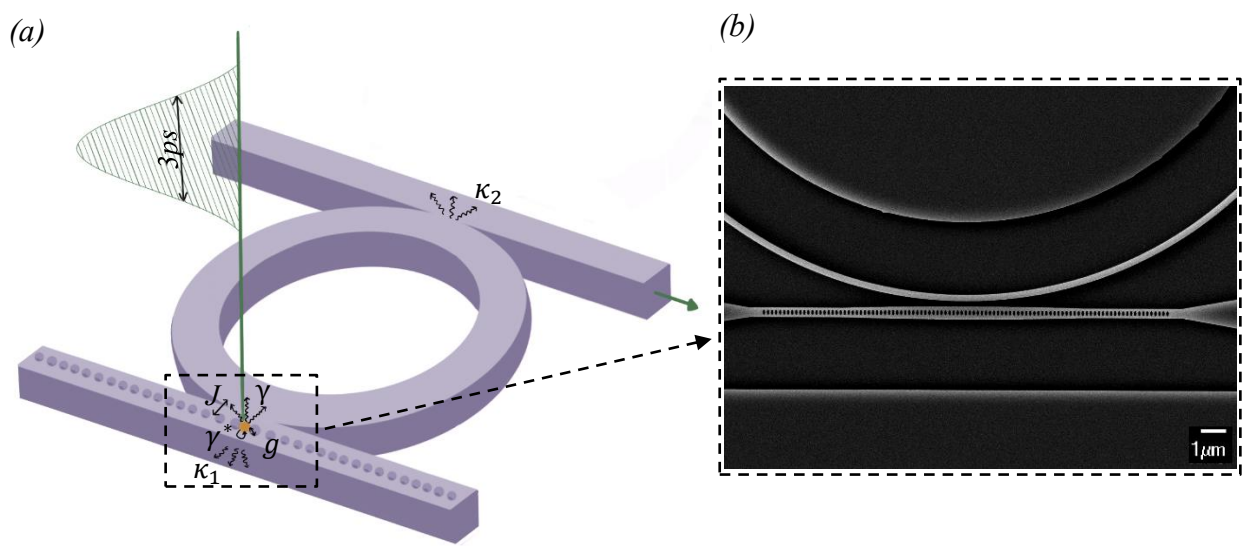

*Figure 5: Experimental design. (a) Design schematic depicting colloidal QD with decay rate γ and dephasing rate γ\* coupled to the nanobeam cavity with a coupling rate of g. The nanobeam*

cavity has a decay rate $\kappa_2$ and is coupled to a ring resonator with a coupling rate J. The ring resonator has a decay rate of $\kappa_2$. The QD is excited by 3ps wide pulse with an amplitude of $P_o = 120\gamma$. The output is collected from the waveguide coupled to the ring resonator. (b) SEM image of a fabricated device structure inside the dotted black box shown in (a).

As shown in the Figure 5(a), our system consists of a colloidal QD coupled to SiN nanobeam cavity which is further coupled to a SiN ring resonator. The colloidal QD is characterized[10] by its decay time constant of $\tau = 4.8ns$ and linewidth of $\Delta\lambda = 23nm$ which can be used to find its decay rate $\gamma$ and dephasing rate $\gamma^*$.

$$\gamma = \frac{1}{\tau}, \qquad \gamma^* = \Delta\omega - \gamma, \qquad \Delta\omega = \frac{\omega_o^2 \Delta\lambda}{2\pi c}$$

where $\omega_o$ is the emission frequency equivalent to $\lambda = 630nm$. The QD is coupled to the nanobeam cavity with a coupling rate of $g$ which depends on the mode volume of the first cavity and the dipole moment of the QD and is given by[23,24]

$$g = \eta \sqrt{\frac{\mu^2 \omega_o}{2\hbar\epsilon_{SiN}\epsilon_o V_{eff}}}$$

where $\eta$ is the relative strength of electric field at colloidal QD location ($E_{CQD}/E_{max}$) and is $= 0.35$ in our case[10] as the QD sits on the surface of the cavity, $\mu$ denotes the dipole moment of the colloidal QD and $\approx 50D$ for the QD used[25]. The nanobeam cavity has a decay rate of $\kappa_1 = \omega_o/Q_1$ and is coupled to the ring resonator with a coupling rate $J = 2.1\gamma$. The ring resonator decays at a rate of $\kappa_2 = \omega_o/Q_2$. The QD radiates at $\lambda = 630nm$ and cavities are designed to have zero detuning. We pump the quantum dot with $3ps$ pulse which has an amplitude $P_o = 120\gamma$. Figure 5(b) shows a scanning electron microscope (SEM) of a fabricated concept device to highlight that such heterogenous integration of different cavities is indeed possible.

Via numerically simulating the master equation using these parameters, we plot the population in the colloidal QD, the nanobeam and the ring resonator as a function of time (Figure 6). Initially all three are in the ground state. As the pulse excites the emitter the population of QD rapidly rises before it starts to drop. Due to the coupling between the QD and the nanobeam, and the coupling between the nanobeam and the ring resonator, population in the cavities too show a similar behavior but with a decreasing magnitude of peak population.

A key difference to note here from the case where we start with an initially excited emitter is that, since there exists a finite period of rise of populations, the net transfer to the second cavity (ring resonator) is lower in this case and we get a lower efficiency though indistinguishability isn't significantly affected. We achieve an efficiency of 0.152% and indistinguishability of 0.629 using our system.

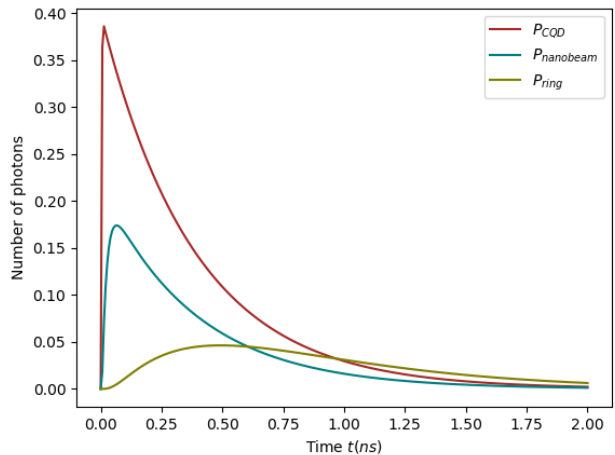

*Figure 6: Population dynamics. We plot how the population of photons in the colloidal quantum dot, the nanobeam and ring resonator given by $P_{CQD}, P_{nanobeam}$ & $P_{ring}$ respectively change with time. Population of the nanobeam has been multiplied by a factor 20 and the population of the ring resonator has been multiplied by a factor of 50. It is evident that such a process isn't very efficient, but it ensures a high indistinguishability of photons collected from the ring resonator.*

**Comparison with other solid-state emitters**

Finally, we compare the performance of colloidal QD as a source of indistinguishable single photons with other quantum emitters like self-assembled QDs and SiV centers. As seen from Table 1 colloidal QDs suffer from a dephasing rate which is almost two orders of magnitude greater than SiV centers and about three orders of magnitude greater than self-assembled QDs. The first column lists the indistinguishability and efficiency of a dissipative self-assembled QD coupled to a single cavity. Indistinguishability of such a system depends on the detuning between the cavity and the emitter and falls with an increase in detuning. The second column details the performance of a single SiV center coupled to system of cascaded cavities for two different parameter values. We can see that an increased indistinguishability comes at a cost of lower efficiency of emission. In the third column we list the performance of colloidal QD coupled to our proposed system of two coupled cavities at an optimal point of operation. As evident, despite the huge amount of dephasing present in solution-processed colloidal QDs we can still get comparable indistinguishability and efficiency from these. However, as explained in the last section, current fabrication techniques pose limitations on the system parameters that can be presently achieved on a nanophotonic platform. Hence, in the last column we list the performance of a colloidal QD coupled to our system of two coupled cavities characterized by parameters currently within the reach of experimental techniques. We can see that SiV centers yield high values of indistinguishability of emitted photons when coupled to a cavity with extremely low mode volume of the order of $\sim 0.005(\lambda/n)^3$. The caveat however is that such small mode volume cavities were only demonstrated on a high refractive index material platform, which is partially absorptive at the SiV resonance frequency.

*Table 1: Comparison between performance of broad quantum emitters: self-assembled QDs, SiV centers and colloidal QDs, as sources of indistinguishable single photons under incoherent pumping. Performance of SiV center has been listed for two sets of parameters taken from the cited paper. $V_{effective}$ in the third row is the mode volume of the cavity to which the emitter is coupled. Note: results used for self-assembled QDs and SiV centers from the referenced papers have been updated to include the effect of pulsed incoherent pumping.*

| *Category* | *Self-assembled QD in a single cavity*[13,26] | *SiV center in coupled cavities*[11] | *Colloidal QD in coupled cavities (optimal)* | *Colloidal QD in coupled cavities (experimental)* |
|---|---|---|---|---|
| $\gamma^*/\gamma$ | 117 | 2500 | 83000 | 83000 |
| $Q_1$ & $Q_2$ | ~$5\times10^4$ | $7\times10^3$ & $5\times10^5$ / $3.6\times10^3$ & $5\times10^4$ | $6\times10^4$ & $2\times10^6$ | $6\times10^4$ & $2\times10^6$ |
| $V_{effective}$ | ~$(\lambda/n)^3$ | $0.007(\lambda/n)^3$ | $0.1(\lambda/n)^3$ | $1.2(\lambda/n)^3$ |
| *Indistinguishability* | ~0.6 | 0.94/0.78 | 0.9 | 0.63 |
| *Efficiency* | 12.1% | 0.26%/0.99% | 0.24% | 0.15% |

**Conclusion**

We proposed an experimentally feasible system design to improve indistinguishability of single photons from colloidal QDs. We looked at the trends in indistinguishability and efficiency of emitted photons as a function of system parameters. Using qualitative arguments, we provided a physical insight on how the system functions. Finally, we compared the performance of colloidal QDs with other broad quantum emitters. We found performance of colloidal QDs to be comparable to these other emitters. Even better performance from colloidal QD as a source of indistinguishable single photons is within reach in near future by using lower mode volume SiN cavities to couple to the colloidal QD. Another viable alternate is to use gallium phosphide (GaP) cavities, as GaP has a much higher refractive index (n = 3.25). This high refractive index contrast allows significantly lower mode volumes of the order of ~ $0.1(\lambda/n)^3$ to be achieved on a GaP based nanophotonic platform. As shown in parameter sweeps and Table 1. using such low mode volumes in either platform, we can achieve indistinguishability greater than 0.9 using colloidal QDs. This number is expected to improve further, thanks to progress in synthesis techniques and new materials like Perovskite QDs[27]. Our work lays a solid foundation for obtaining indistinguishable photons from colloidal QDs coupled to a nanophotonic platform and can potentially solve the long-standing challenge of the scalable quantum photonic technology.

**Acknowledgement:** This work is supported by NSF Award 1836500. A.S. is supported by a CEI graduate fellowship. A.R. is supported by the IC Postdoctoral Fellowship.

# Supplementary Material: Improving indistinguishability of single photons from colloidal quantum dots using nanocavities

*Abhi Saxena[1], Yueyang Chen[1], Albert Ryou[1], Carlos G. Sevilla[3], Peipeng Xu[1,4,5], Arka Majumdar[1,2,*]*

[1]*Electrical and Computer Engineering, University of Washington, Seattle, Washington 98195, United States*

[2]*Department of Physics, University of Washington, Seattle, Washington 98195, United States*

[3]*School of Natural Science, Hampshire College, Amherst, MA 01002, United States*

[4]*Laboratory of Infrared Materials and Devices, Advanced Technology Research Institute, Ningbo University, Ningbo 315211, China*

[5]*Key Laboratory of Photoelectric Detection Materials and Devices of Zhejiang Province, Ningbo, 315211, China*

[*]*Corresponding Author:* arka@uw.edu

## S1. Mathematical Framework:

We use the master equation and the optical Bloch equations to model our system of a quantum emitter coupled to a system of two nanophotonic resonators.[1–3] Our proposed system consists of a quantum emitter coupled to a nanophotonic cavity $C_1$ with a coupling rate $g$. The emitter has a decay rate of $\gamma$ and suffers from dephasing at a rate $\gamma^*$. The cavity $C_1$ has a decay rate of $\kappa_1$. $C_1$ is coupled to another cavity $C_2$ with a coupling rate $J$. $C_2$ is characterized by a decay rate of $\kappa_2$.

The system is governed by the Hamiltonian (setting $\hbar = 1$)

$$H = \omega_e e^\dagger e + \omega_{c_1} c_1^\dagger c_1 + \omega_{c_2} c_2^\dagger c_2 + g(e^\dagger c_1 + e c_1^\dagger) + J(c_1^\dagger c_2 + c_1 c_2^\dagger) \quad (1)$$

where $e^\dagger, c_1^\dagger, c_2^\dagger$ are the creation operators for the emitter and the cavities $C_1$ and $C_2$ respectively. As there is only one quantum of energy, the state space of the system consists of $|0,0,0\rangle, |1,0,0\rangle, |0,1,0\rangle, |0,0,1\rangle$ where the first index denotes the emitter, the second denotes $C_1$ and the last denotes $C_2$. Depending on the index, '0' denotes the emitter in ground state or the cavities being empty, '1' denotes an excited emitter or the photon number of the filled cavities. In this state space Hamiltonian (using rotating wave approximation) is written as

$$H = \begin{bmatrix} 0 & 0 & 0 & 0 \\ 0 & 0 & g & 0 \\ 0 & g & 0 & J \\ 0 & 0 & J & 0 \end{bmatrix} \quad (2)$$

The system dynamics is given by the evolution of the density matrix according to the master equation

$$\frac{\partial \rho}{\partial t} = -i[H, \rho(t)] + \sum_n \left[ \frac{1}{2} \left( 2A_n \rho(t) A_n^\dagger - \rho(t) A_n^\dagger A_n - A_n^\dagger A_n \rho(t) \right) \right] \quad (3)$$

where $A_n$ denotes the collapse operators required to model the system: $\sqrt{\kappa_1}c_1, \sqrt{\kappa_2}c_2, \sqrt{\gamma}e, \sqrt{\gamma^*}e^\dagger e, \sqrt{P(t)}e^\dagger$ where the last term represents incoherent pumping of the QD. In our state space the density matrix $\rho$ can be written as

$$\rho = \begin{bmatrix} \rho_{00} & \rho_{01} & \rho_{0c_1} & \rho_{0c_2} \\ \rho_{10} & \rho_{11} & \rho_{1c_1} & \rho_{1c_2} \\ \rho_{c_10} & \rho_{c_11} & \rho_{c_1c_1} & \rho_{c_1c_2} \\ \rho_{c_20} & \rho_{c_21} & \rho_{c_2c_1} & \rho_{c_2c_2} \end{bmatrix} \quad (4)$$

where $\rho_{00}, \rho_{11}, \rho_{c_1c_1}, \rho_{c_2c_2}$ denote the population in the ground state of the emitter, the population in the excited state of the emitter, the photon population in $C_1$ and the photon population in $C_2$ correspondingly. Terms of the type $\rho_{xy}$ denote coherences between state 'x' and 'y'. Of course, $\rho_{xy} = \rho_{yx}^*$. Hence there are only seven independent variables in the density matrix. In this state space, the master equation can be expanded to give a system of relevant differential equations, the optical Bloch equations which describe the system population dynamics. These are

$$\frac{\partial \rho_{00}}{\partial t} = -P(t)\rho_{00} + \gamma \rho_{11} + \kappa_1 \rho_{c_1c_1} + \kappa_2 \rho_{c_2c_2} \quad (5)$$

$$\frac{\partial \rho_{11}}{\partial t} = P(t)\rho_{00} - \gamma \rho_{11} + ig(\rho_{1c_1} - \rho_{c_11}) \quad (6)$$

$$\frac{\partial \rho_{c_1c_1}}{\partial t} = -\kappa_1 \rho_{c_1c_1} + ig(\rho_{c_11} - \rho_{1c_1}) + iJ(\rho_{c_1c_2} - \rho_{c_2c_1}) \quad (7)$$

$$\frac{\partial \rho_{c_2c_2}}{\partial t} = -\kappa_2 \rho_{c_2c_2} + iJ(\rho_{c_2c_1} - \rho_{c_1c_2}) \quad (8)$$

$$\frac{\partial \rho_{1c_1}}{\partial t} = -\frac{\gamma + \gamma^* + \kappa_1}{2}\rho_{1c_1} + ig(\rho_{11} - \rho_{c_1c_1}) + iJ\rho_{1c_2} \quad (9)$$

$$\frac{\partial \rho_{1c_2}}{\partial t} = -\frac{\gamma + \gamma^* + \kappa_2}{2}\rho_{1c_2} + i(J\rho_{1c_1} - g\rho_{c_1c_2}) \quad (10)$$

$$\frac{\partial \rho_{c_1c_2}}{\partial t} = -\frac{\kappa_1 + \kappa_2}{2}\rho_{c_1c_2} - ig\rho_{1c_2} + iJ(\rho_{c_1c_1} - \rho_{c_2c_2}) \quad (11)$$

In our system the quantum emitter is highly dissipative due to a large dephasing rate. The dephasing rate of the emitter $\gamma^*$ is much larger than other decay rates of the system $\gamma, \kappa_1, \kappa_2$. Thus, for a timescale $t > 1/\gamma^*$ when $\gamma + \gamma^* + \kappa_2 \gg 2J$ or $2g$ coherence between the emitter and cavity $C_2$ can be eliminated ($\frac{\partial \rho_{1c_2}}{\partial t} \approx 0$) to obtain.

$$\rho_{1c_2} = \frac{2i(J\rho_{1c_1} - g\rho_{c_1 c_2})}{\gamma + \gamma^* + \kappa_2} \tag{12}$$

Substituting Eq. 12 into Eq. 9 and Eq. 11 gives us

$$\frac{\partial \rho_{1c_1}}{\partial t} = -\frac{\gamma + \gamma^* + \kappa_1}{2}\rho_{1c_1} + ig(\rho_{11} - \rho_{c_1 c_1}) - \frac{2J(J\rho_{1c_1} - g\rho_{c_1 c_2})}{\gamma + \gamma^* + \kappa_2} \tag{13}$$

$$\frac{\partial \rho_{c_1 c_2}}{\partial t} = -\frac{\kappa_1 + \kappa_2}{2}\rho_{c_1 c_2} + iJ(\rho_{c_1 c_1} - \rho_{c_2 c_2}) + \frac{2g(J\rho_{1c_1} - g\rho_{c_1 c_2})}{\gamma + \gamma^* + \kappa_2} \tag{14}$$

Rearranging the terms from Eq. 14, we get

$$\frac{\partial \rho_{c_1 c_2}}{\partial t} = -\frac{\kappa_1 + \kappa_2 + \frac{4g^2}{\gamma + \gamma^* + \kappa_2}}{2}\rho_{c_1 c_2} + iJ(\rho_{c_1 c_1} - \rho_{c_2 c_2}) + \frac{2gJ\rho_{1c_1}}{\gamma + \gamma^* + \kappa_2}$$

We now define a population transfer rate $R_1$ between emitter and $C_1$ as

$$R_1 = \frac{4g^2}{\gamma + \gamma^* + \kappa_1} \tag{15}$$

Hence, we get

$$\frac{\partial \rho_{c_1 c_2}}{\partial t} = -\frac{\kappa_1 + \kappa_2 + R_1\frac{\gamma + \gamma^* + \kappa_1}{\gamma + \gamma^* + \kappa_2}}{2}\rho_{c_1 c_2} + iJ(\rho_{c_1 c_1} - \rho_{c_2 c_2}) + \frac{2gJ\rho_{1c_1}}{\gamma + \gamma^* + \kappa_2}$$

The term $\frac{2gJ\rho_{1c_1}}{\gamma+\gamma^*+\kappa_2}$ can be neglected because $J\rho_{1c_1} \ll g\rho_{c_1c_2}$ for $t > 1/R_1$. Also, as $\gamma^* \gg \kappa_1, \kappa_2, \gamma$ we can assume $\frac{\gamma+\gamma^*+\kappa_1}{\gamma+\gamma^*+\kappa_2} \approx 1$. For $t > 1/R_1$, $\frac{\partial \rho_{c_1c_2}}{\partial t}$ can be adiabatically eliminated to give

$$\rho_{c_1c_2} = \frac{2iJ(\rho_{c_1c_1} - \rho_{c_2c_2})}{\kappa_1 + \kappa_2 + R_1} \tag{16}$$

Substituting Eq. 16 into Eq. 13 we get

$$\frac{\partial \rho_{1c_1}}{\partial t} = -\frac{\gamma + \gamma^* + \kappa_1 + \frac{4J^2}{\gamma+\gamma^*+\kappa_2}}{2}\rho_{1c_1} + ig(\rho_{11} - \rho_{c_1c_1}) + \frac{4igJ^2(\rho_{c_1c_1} - \rho_{c_2c_2})}{(\gamma+\gamma^*+\kappa_2)(\kappa_1+\kappa_2+R_1)} \tag{17}$$

Further, ignoring $\Theta(1/\gamma^*)$ terms gives us

$$\frac{\partial \rho_{1c_1}}{\partial t} \approx \frac{\gamma + \gamma^* + \kappa_1}{2}\rho_{1c_1} + ig(\rho_{11} - \rho_{c_1c_1}) \tag{18}$$

Finally, by adiabatic elimination for $t > 1/\gamma^*$

$$\rho_{1c_1} = \frac{2ig(\rho_{11} - \rho_{c_1c_1})}{\gamma + \gamma^* + \kappa_1} \tag{19}$$

At this stage from Eq. 12, Eq. 16 and Eq. 19 we can see that the coherences of the system follow the populations under the specified conditions on the system. Thus, the population dynamics given by rate equations completely describe the system. These can be obtained by substituting Eq. 12, Eq. 16 and Eq. 19 into Eq. 5, Eq. 6, Eq. 7 and Eq. 8.

$$\frac{\partial \rho_{00}}{\partial t} = -P(t)\rho_{00} + \gamma\rho_{11} + \kappa_1\rho_{c_1c_1} + \kappa_2\rho_{c_2c_2}$$

$$\frac{\partial \rho_{11}}{\partial t} = P(t)\rho_{00} - \gamma\rho_{11} - \frac{4g^2}{\gamma + \gamma^* + \kappa_1}(\rho_{11} - \rho_{c_1 c_1}) \qquad (20)$$

$$\frac{\partial \rho_{c_1 c_1}}{\partial t} = -\kappa_1 \rho_{c_1 c_1} - \frac{4g^2}{\gamma + \gamma^* + \kappa_1}(\rho_{c_1 c_1} - \rho_{11}) - \frac{4J^2}{\kappa_1 + \kappa_2 + R_1}(\rho_{c_1 c_1} - \rho_{c_2 c_2}) \qquad (21)$$

$$\frac{\partial \rho_{c_2 c_2}}{\partial t} = \kappa_2 \rho_{c_2 c_2} - \frac{4J^2}{\kappa_1 + \kappa_2 + R_1}(\rho_{c_2 c_2} - \rho_{c_1 c_1}) \qquad (22)$$

Now we define $R_2$, the population transfer rate between cavities $C_1$ and $C_2$, $R_2 = \frac{4J^2}{\kappa_1 + \kappa_2 + R_1}$. Substituting this into the above equations gives us,

$$\frac{\partial \rho_{00}}{\partial t} = -P(t)\rho_{00} + \gamma\rho_{11} + \kappa_1 \rho_{c_1 c_1} + \kappa_2 \rho_{c_2 c_2}$$

$$\frac{\partial \rho_{11}}{\partial t} = P(t)\rho_{00} - (\gamma + R_1)\rho_{11} + R_1 \rho_{c_1 c_1} \qquad (23)$$

$$\frac{\partial \rho_{c_1 c_1}}{\partial t} = R_1 \rho_{11} - (\kappa_1 + R_1 + R_2)\rho_{c_1 c_1} + R_2 \rho_{c_2 c_2} \qquad (24)$$

$$\frac{\partial \rho_{c_2 c_2}}{\partial t} = R_2 \rho_{c_1 c_1} - (\kappa_2 + R_2)\rho_{c_2 c_2} \qquad (25)$$

Hence, we see that the system dynamics can be completely described by the decay rates $\gamma$, $\kappa_1$, $\kappa_2$ and the population transfer rates $R_1$ and $R_2$ as depicted in Figure 2 in the paper.

## S2. Nanobeam cavity

The cavity mode profile is shown in Figure S1. The simulated SiN nanobeam cavity has a resonant wavelength at $630 nm$, quality factor $Q_1 = 6 \times 10^4$ and a mode volume of $V_{eff} = 1.2(\lambda/n)^3$.

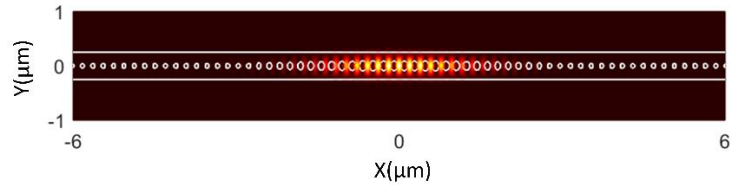

*Figure S1: Mode profile ( $|E|^2$) of the SiN nanobeam photonic crystal cavity*

The design parameters are as follow:

The SiN nanobeam is sitting on a silicon oxide substrate without any cladding on top. The nanobeam has a thickness t = 220 nm and a width w = 490 nm. The periodicity of the elliptical holes is fixed as 191nm. The Bragg region consists of 45 elliptical holes on each side, with a major and a minor diameter of 98 nm and 67 nm, respectively. Both the major diameter and minor diameter are quadratically tapered from the Bragg region to the center of the nanobeam. The tapering region consists of 15 holes on each side. The major diameter and minor diameter of the elliptical hole that is closest to the center of the nanobeam is 176 nm and 115nm, respectively.